\documentclass[twocolumn,showpacs,preprintnumbers,amsmath,amssymb,
superscriptaddress]{revtex4}

\usepackage{graphicx}
\usepackage{dcolumn}
\usepackage{bm}
\usepackage{xcolor}

\definecolor{lightgray}{gray}{0.9}

\begin{document}
	
\preprint{}
	
\title{Gapped Excitations in the High-Pressure Antiferromagnetic Phase of URu$_2$Si$_2$}
	
\author{T.J.~Williams}
\email{williamstj@ornl.gov}
\affiliation{Department of Physics and Astronomy,
	McMaster University, Hamilton, ON, Canada, L8S 4M1}
\affiliation{Quantum Condensed Matter Division,
	Neutron Sciences Directorate,
	Oak Ridge National Lab,
	Oak Ridge, TN, 37831, USA}

\author{H.~Barath}
\affiliation{Institute for Quantum Matter and Department
	of Physics and Astronomy, Johns Hopkins University,
	Baltimore, MD, USA, 21218}

\author{Z.~Yamani}
\affiliation{National Research Council,
	Chalk River Laboratories,
	Chalk River, ON, Canada, K0J 1J0}

\author{J.A.~Rodriguez-Riviera}
\affiliation{NIST Center for Neutron Research, National 
	Institute of Standards and Technology, Gaithersburg, MD,
	USA, 20899}
\affiliation{Department of Materials Science and Engineering, 
	University of Maryland, College Park, MD, USA, 20740}

\author{J.B.~Le\~{a}o}
\affiliation{NIST Center for Neutron Research, National 
	Institute of Standards and Technology, Gaithersburg, MD,
	USA, 20899}

\author{J.D.~Garrett}
\affiliation{Brockhouse Institute for Materials Science,
	McMaster University, Hamilton, ON, Canada, L8S 4M1}

\author{G.M. Luke}
\affiliation{Department of Physics and Astronomy,
	McMaster University, Hamilton, ON, Canada, L8S 4M1}
\affiliation{Brockhouse Institute for Materials Science,
	McMaster University, Hamilton, ON, Canada, L8S 4M1}
\affiliation{Canadian Institute for Advanced Research,
	Toronto, Ontario, Canada, M5G 1Z8}

\author{W.J.L.~Buyers}
\affiliation{National Research Council,
	Chalk River Laboratories,
	Chalk River, ON, Canada, K0J 1J0}
\affiliation{Canadian Institute for Advanced Research,
	Toronto, Ontario, Canada, M5G 1Z8}

\author{C.~Broholm}
\email{broholm@jhu.edu}
\affiliation{Institute for Quantum Matter and Department
	of Physics and Astronomy, Johns Hopkins University,
	Baltimore, MD, USA, 21218}
\affiliation{NIST Center for Neutron Research, National 
	Institute of Standards and Technology, Gaithersburg, MD,
	USA, 20899}

\date{\today}

\begin{abstract}
We report a neutron scattering study of the magnetic excitation spectrum 
in each of the three temperature and pressure driven phases of URu$_2$Si$_2$. 
We find qualitatively similar excitations throughout the (H0L) scattering 
plane in the hidden order and large moment phases, with no changes in the 
$\hbar\omega$-widths of the excitations at the $\Sigma$~=~(1.407,0,0) and 
$Z$~=~(1,0,0) points, within our experimental resolution.  There is, however, 
an increase in the gap at the $\Sigma$ point from 4.2(2)~meV to 5.5(3)~meV, 
consistent with other indicators of enhanced antiferromagnetism under 
pressure. 
\end{abstract}

\pacs{78.70.Nx, 71.27.+a}

\maketitle

The heavy fermion material URu$_2$Si$_2$ exhibits a specific heat anomaly at 
T$_0$~=~17.5~K indicative of a second order phase 
transition~\cite{Palstra_85,Broholm_87}.  Decades of research not 
withstanding~\cite{Bonn_88,Buyers_94,Mason_95,Bourdarot_05}, an order 
parameter characterizing the putative symmetry breaking of the low temperature 
phase has not been identified.  Neutron scattering does show antiferromagnetic 
order with an ordering wavevector ${\bf Q}_m$~=~(1,0,0), but the small 
sample-averaged moment of 0.03~$\mu_B$~\cite{Broholm_87}, seems hard to 
reconcile with a change in entropy $\Delta S$~=~0.24$R \ln 
2$~\cite{Palstra_85} through the transition.  This moment may even be 
intrinsic~\cite{Bourdarot_14} or it may arise from heterogeneous inclusions of 
a large moment phase~\cite{Takagi_07}.  Spin fluctuations with a 
characteristic wave vector (1~$\pm$~$\delta$,~0,~0) ($\delta$~=~0.407(6)) are 
observed in the paramagnetic (PM) phase, indicative of Fermi-surface nesting 
at the $\Sigma$ point, which for URu$_2$Si$_2$ occurs for 
$\delta~=~\frac{1}{2}(1-(a/c)^2)~=~0.406$ (a=4.128~\AA~and c=9.534~\AA~at 
T=4~K)~\cite{Wiebe_07}.  Below $T_0$, in the so-called `hidden order' (HO) 
phase, these excitations become gapped as for a spin density wave transition 
and consistent with the specific heat anomaly, but without development of the 
attendant staggered magnetization.

Hydrostatic pressure of $\sim$0.6~GPa replaces the HO phase with a 
large-moment antiferromagnetic (AF) phase with an ordered magnetic moment of 
0.3~$\mu_B$~\cite{Aoki_09,Butch_10}. Here we show the gapped excitations at 
the $\Sigma$~=~(1.407,0,0) and $Z$~=~(1,0,0) points persist in the AF phase, 
albeit with an enhanced gap at the $\Sigma$ point in the high pressure phase. 
Our results are not inconsistent with previous experimental 
data~\cite{Aoki_09,Hassinger_10a,Bourdarot_10}, though they clearly 
show an inelastic signal at the $Z$ point.  Our expanded coverage of ${\bf 
Q}$-$E$ space reveals a similarity between magnetic excitations in the two 
low temperature phases that was not previously appreciated.

High quality single crystals of URu$_2$Si$_2$ were grown by the 
Czochralski method in a tri-arc furnace.  Three crystals with a total 
mass of approximately 37~g and an RRR~$\approx$~10 were coaligned in the 
($H$0$L$) plane for the ambient pressure measurements.  A single crystal with 
a mass of 1.66~g and an RRR~=~15 was cut by spark erosion, aligned in the 
($H$0$L$) plane, placed inside a 13-8Mo steel He-gas pressure vessel, and 
connected to a commercially available pressurizing intensifier through a 
heated high-pressure capillary. Following the procedure established in 
Ref.~\cite{Butch_10}, the pressure was adjusted only at temperatures 
well above the helium melting curve and the capillary was heated during slow 
cooling of the cell to accommodate the contracting He gas, thus minimizing 
pressure loss and pressure inhomogeneities across the sample space. The 
pressure cell was cooled at constant pressure to the freezing 
point of helium. Through prior calibration measurements of the lattice 
parameters of highly oriented pyrolytic graphite crystals within the cell, the 
pressure reduction upon cooling following these procedures is less than 0.05 
GPa.  The neutron scattering measurements were performed on the Multi-Axis 
Crystal Spectrometer (MACS) at the NIST Center for Neutron Research, where a 
20~MW reactor, a dedicated liquid H$_2$ moderator, and a doubly-focusing 
PG(002) monochromator provides an incident beam flux of 
3.0~$\times$~10$^8$~n/cm$^2$/s~\cite{Rodriguez_08} for an initial energy 
$E_i$~=~5~meV.  In the vicinity of (1,0,0), the in-plane resolution was 
0.12~$\AA^{-1}$ along $L$, 0.043~$\AA^{-1}$ along $H$, and the out-of-plane 
resolution was 0.24~$\AA^{-1}$ at zero energy transfer.  All measurements were 
performed using a fixed $E_f$~=~5.054~meV, with an elastic energy resolution 
of 0.45~meV.  Twenty detection channels permitted efficient mapping of 
inelastic scattering throughout the ($H$0$L$) plane.  Measurements were 
performed at ambient pressure and $T$~=~25~K in the paramagnetic phase, at 
ambient pressure and $T$~=~2~K in the hidden order phase, and at a pressure of 
$P$~=~1.02~GPa and $T$~=~4~K in the AF phase.

\begin{figure}[thb]
\begin{center}
\includegraphics[angle=0,width=\columnwidth]{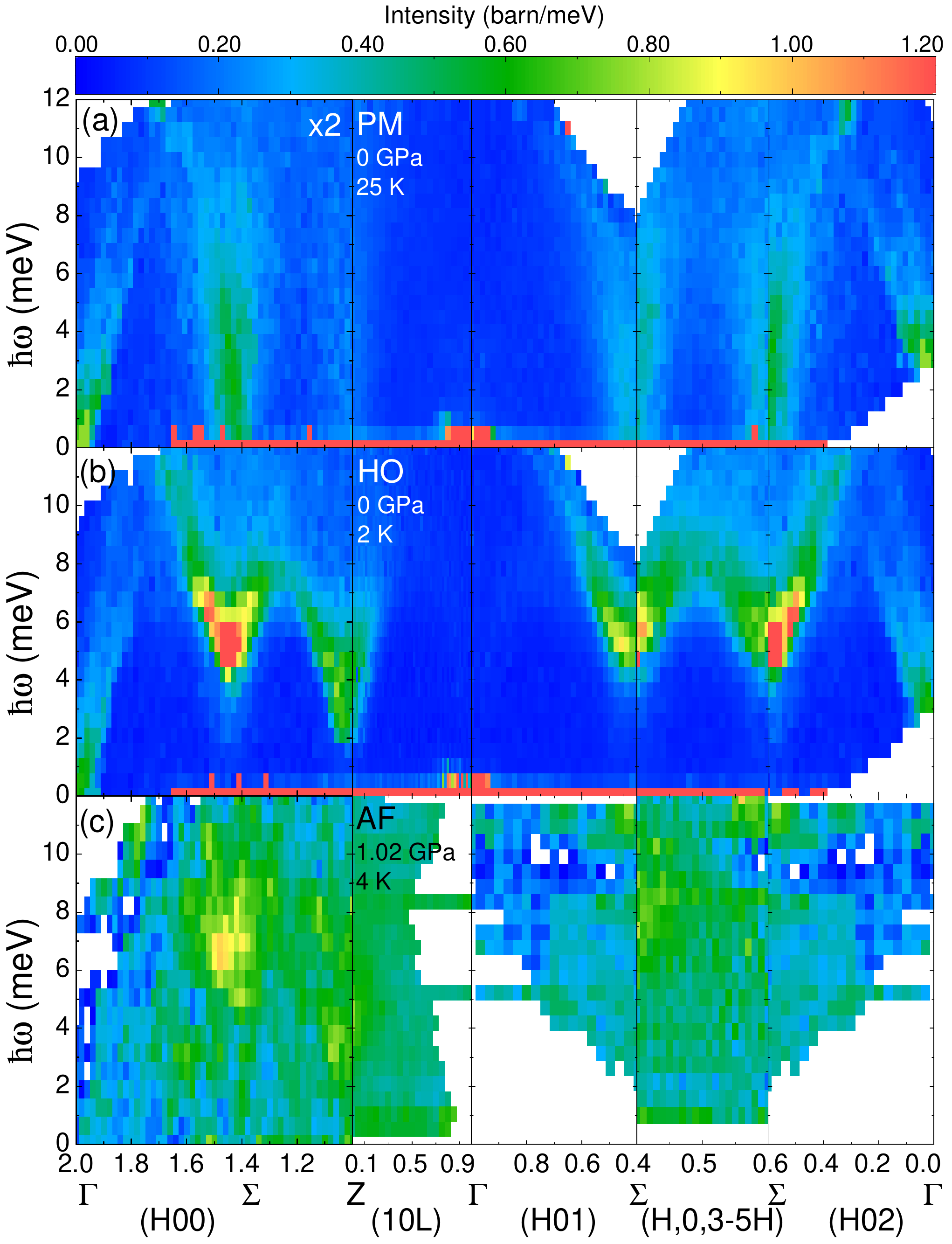}
\caption{\label{hw_vs_H00}
(Color online)
The scattering intensity as a function of energy and scattering vector 
along various high symmetry directions in the ($H$0$L$) plane in the three 
phases studied: (a) at ambient pressure and 25K, in the paramagnetic (PM) 
phase.  For ease of viewing, the data has been scaled up by a factor of 2.  
The phonon at (2,0,0) is visible, as are magnetic excitations at the 
$\Sigma$ point. (b) Data collected at ambient pressure and 2~K, in the HO 
phase. (c) Data collected at 1.02~GPa and 4~K, in the AF phase.  The 
scattering here looks qualitatively similar to the HO phase, albeit with a 
larger gap at the $\Sigma$ point. The reduced statistical quality results from 
the reduced neutron transmission through the pressure cell and the subtraction 
of a strong background signal from the pressure cell and the solid helium 
pressure transmitting medium. }
\end{center}
\end{figure}

Phonon scattering near (0,0,2) is visible in both the paramagnetic 
(Fig.~\ref{hw_vs_H00}(a)) and hidden order (Fig.~\ref{hw_vs_H00}(b)) 
phases.  This allowed for normalization of the data so that we can provide 
absolute values of the scattering cross sections in each phase, which are 
consistent with previously published values~\cite{Butch_15}.  To isolate 
scattering from URu$_2$Si$_2$ from that associated with the massive pressure 
cell and the helium pressure medium, a background was measured for the 
pressure cell with the sample exchanged by an equal volume of aluminum 
pressurized to 1.02~GPa.  Due to the reduced neutron absorption of Al relative 
to URu$_2$Si$_2$, this results in a slight over-subtraction and thus a 
difference signal with a small negative background value, as shown below.  The 
scattering intensity under pressure was also subject to normalization using 
the (0,0,2) structural Bragg peak, which showed the transmission of the 
pressure cell is 18\%, consistent with direct measurements.  Attributing all 
of the scattering at (1,0,0) to magnetic scattering, the normalization yields a cross section for the (1,0,0) magnetic Bragg peak of 0.36(9)~$\mu_B$, which is in good agreement with the previously reported ordered moment in the AF phase~\cite{Aoki_09}.  All data were corrected for the effects of higher order contamination on the monitor count rate~\cite{Rodriguez_08}. 

The inelastic scattering cross section along high symmetry directions in the 
($H$0$L$) plane for the three different phases is shown in 
Fig.~\ref{hw_vs_H00}.  The upper and middle panels show the momentum and 
energy transfer dependence of the magnetic scattering in the PM and HO phases, 
respectively, which are consistent with earlier 
findings~\cite{Broholm_87,Wiebe_07,Wiebe_04}.  There are substantial changes 
across the PM to HO phase transition.  In the PM phase, the scattering takes 
the form of gapless ridges with most of the intensity at the $\Sigma$ point 
though a ridge is also clearly discerned at the $Z$ point.  In the HO phase, 
well-defined gaps have opened at both the $Z$ and $\Sigma$ points, and the 
intensity at the $Z$ point has increased. The lowest panel shows data in the 
AF phase.  Due to the pressure cell the quality of these data is significantly 
reduced. Nonetheless, to within error in the AF phase, the overall Q-$\omega$ 
dependent scattering is qualitatively similar to that of the HO phase, though 
as seen in constant-Q cuts of the data, the gap at the $\Sigma$ point is 
considerably enhanced.

\begin{figure}[!th]
\begin{center}
\includegraphics[angle=0,width=\columnwidth]{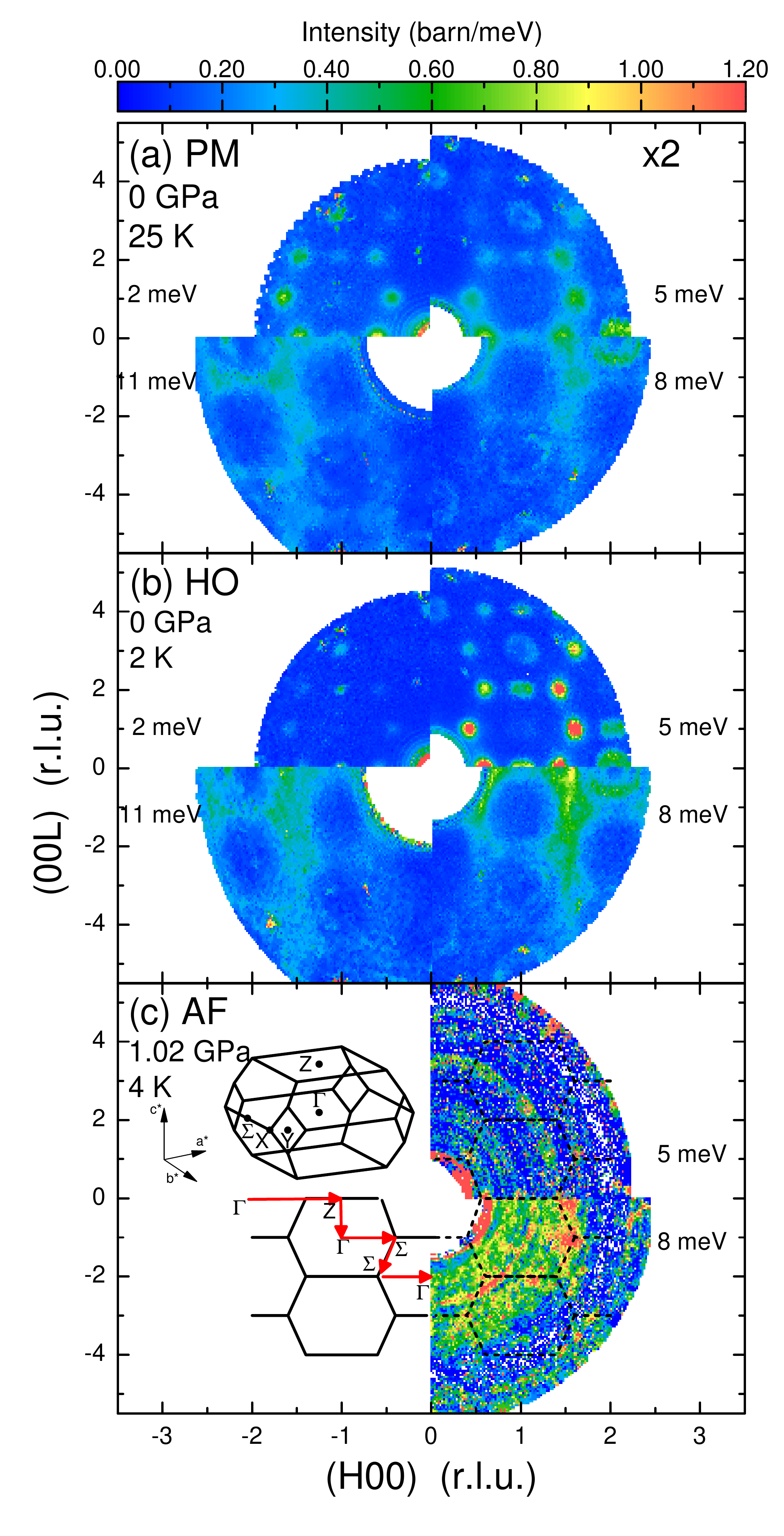}
\caption{\label{slices}
(Color online)
Constant energy slices in the ($H$0$L$) plane in each of the three phases of 
URu$_2$Si$_2$. Energies shown are 2~meV, 5~meV, 8~meV and 11~meV (clockwise 
from top left).  The range of integration of energies for the slices was 
$\pm$0.5~meV and the bin size is 0.013~\AA$^{-2}$.  The data in the PM phase 
(panel (a)) has been scaled up by a factor of 2, as in Fig.~\ref{hw_vs_H00}.  
The lower left of the figure shows the Brillouin zone of URu$_2$Si$_2$, with 
the arrows indicating the directions shown in Fig.~\ref{hw_vs_H00}.
}
\end{center}
\end{figure}

Further comparisons between the three phases is made by examining the 
constant energy transfer slices through the ($H$0$L$) zone data, shown in 
Fig.~\ref{slices}.  The figure shows the average intensity in 1~meV-thick 
slices centered at 2~meV, 5~meV, 8~meV and 11~meV.  For improved statistics, 
we have symmetrized the data and present a single quadrant at each energy 
transfer.  While intensity at the $\Sigma$ point is present in all 
three phases, intensity at the $Z$ point is mainly visible in the HO and AF 
phases.  In the HO phase, there is considerably more spectral weight in all of 
the excitations compared to the PM phase.  Comparing the HO and AF phase, we 
see in the 5~meV slice that the gap in the AF phase is larger at the 
$\Sigma$ points.  The $Z$ and $\Sigma$ modes have similar intensity at 8~meV 
within the AF phase compared to 5~meV in the HO phase.  In the HO phase, the 
8~meV data consists of smooth ridges, while well-defined reciprocal space 
intensity maxima are still visible at 8~meV in the AF data.

\begin{figure}[!th]
\begin{center}
\includegraphics[angle=0,width=\columnwidth]{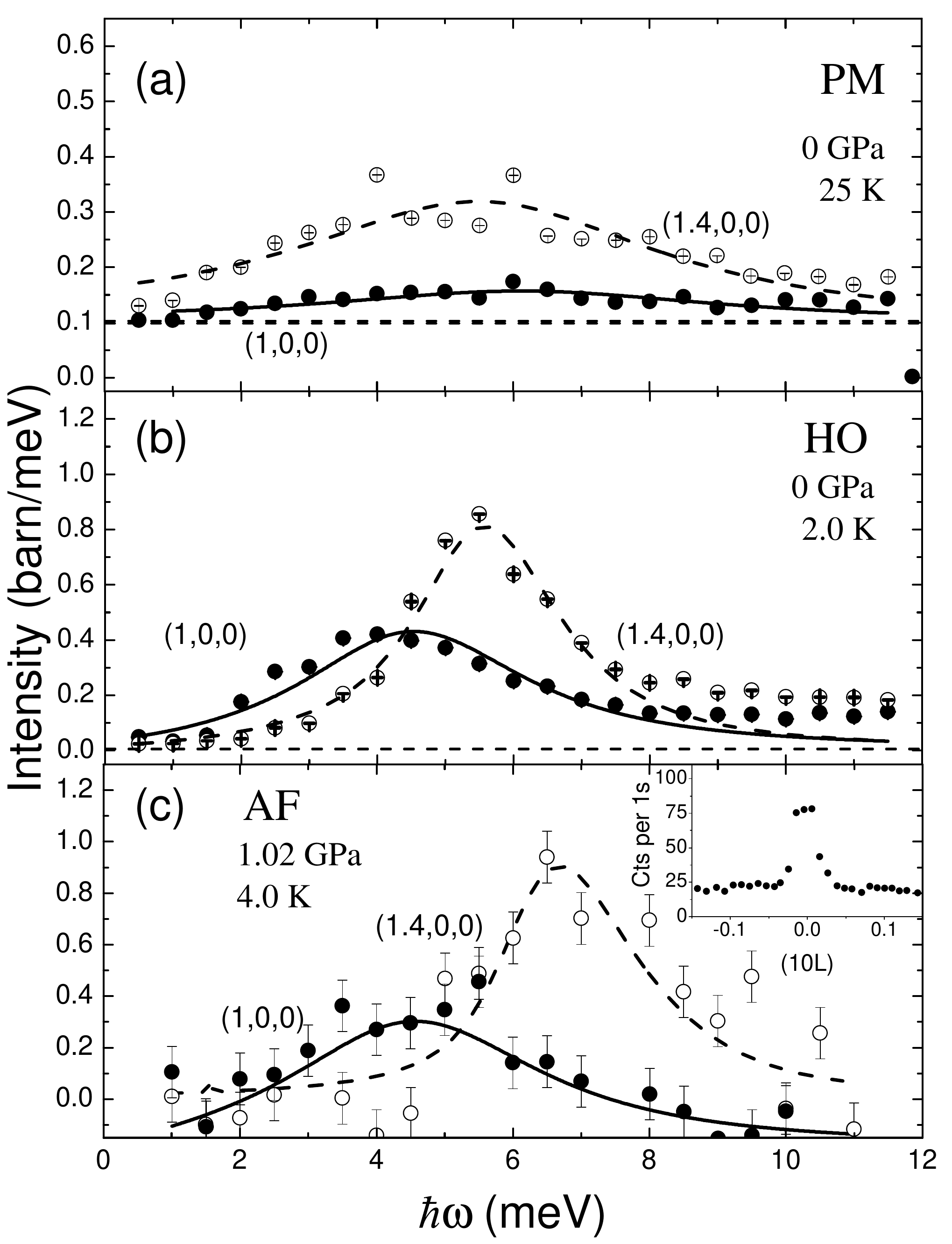}
\caption{\label{I_vs_hw}
(Color online)
The energy-dependence of the scattering intensity for crystal momentum $Z$ 
(filled circles) and $\Sigma$ (open circles) in each phase.  The error bars 
represent 1 standard deviation, $\sigma$.  The lines are fits as described in 
the text, with the horizontal dashed lines showing the fitted background. (a) 
In the PM phase, a weak and broad spectrum of scattering is seen at $Z$, with 
a more pronounced but also broad peak at $\Sigma$. (b) In the HO phase, the 
scattering is more intense at both wave vectors and intensity at $\Sigma$ has 
shifted to higher energy. (c) In the AF phase, both peaks are still present; 
the peak at (1.4,0,0) has shifted to slightly higher energies. Note: the 
fitted background lies below the axis here, due to the subtraction described 
in the text. Inset: the scattering along (10$L$) in the AF phase. When 
normalized, the peak intensity at (1,0,0) corresponds to a sample averaged 
staggered moment of 0.36(9)~$\mu_B$.}
\end{center}
\end{figure}

For a quantitative spectral analysis, the energy dependence of the scattering 
at $Z$ and $\Sigma$ in the three phases is shown in Fig.~\ref{I_vs_hw}.  
These cuts were extracted from the same data that is shown in 
Fig.~\ref{hw_vs_H00} and Fig.~\ref{slices}.  Following the analysis 
of~\cite{Broholm_91}, the data in each phase was fit to the 
resolution-convoluted line shape associated with the following expression for 
the low temperature magnetic scattering cross section near the $Z$ and 
$\Sigma$ points: 

\begin{eqnarray}
\tilde{I}({\bf Q},\omega)=
& \frac{\cal A}{\epsilon({\bf Q})} \cdot
\frac{1-e^{-\beta\Delta}}{1-e^{-\beta\hbar\omega}} \cdot
\nonumber \\
& \left [ \frac{\hbar\gamma/\pi}{(\hbar\omega-\epsilon({\bf Q}))^2
+(\hbar\gamma)^2}
-\frac{\hbar\gamma/\pi}{(\hbar\omega+\epsilon({\bf Q}))^2
+(\hbar\gamma)^2} \right ]
\label{eq1}
\end{eqnarray}

\noindent where $\hbar\gamma$ is the spectral Half Width at Half Maximum 
(HWHM) and ${\cal A} \approx \hbar^2\int\tilde{I}({\bf Q},\omega)\omega 
d\omega$ approximates the first moment in the limit where $\hbar\gamma \ll 
\epsilon({\bf Q})$.  With an energy gap $\Delta$, the phenomenological 
dispersion relation reads:

\begin{equation}
\epsilon({\bf Q})=
\sqrt{\Delta^2+\hbar^2(\delta Q_{\perp}^2v_{\perp}^2+\delta 
Q_{\parallel}^2v_{\parallel}^2)}
\label{eq2}
\end{equation}

Here $\delta Q_{\perp,\parallel}=|({\bf Q}-{\bf Q}_0)_{\perp,\parallel}|$ is 
the projection of the deviation in wave vector transfer ${\bf Q}$ from the 
critical wave vector ${\bf Q}_0$ perpendicular and parallel, respectively, to 
the $\mathbf{\hat{c}}$-direction. We take the velocity to be 
isotropic within the tetragonal basal plane because the present data from the 
($H$0$L$) zone only is insensitive to  potential in-plane anisotropy allowed 
by symmetry in the low $T$ phases.  The velocities used were determined from 
the HO phase, using the data in Fig.~\ref{hw_vs_H00}(b), and were found to be 
$v_H$~=~$v_K$~=~$v_{\perp}$~=~23.7(5)~meV$\cdot$\AA~and 
$v_L$~=~$v_{\parallel}$~=~32.5(7)~meV$\cdot$\AA.  Eq.~\ref{eq1} was  
convoluted with the 4D instrumental resolution function using 
\textsc{Reslib}~\cite{Zheludev_07}.  In order to extract reliable measurements 
of the energy gaps at both ${\bf Q}$-points, this fitting was performed for a 
variety of integration ranges in both $H$ and $L$.  The values of the gap, 
$\Delta$, and width, $\hbar\gamma$, versus the integration area (in 
\AA$^{-2}$) were then extrapolated to the size of the resolution ellipse given 
by \textsc{Reslib}.  This allowed these parameters to be determined in a way 
that is only dependent on the instrumental resolution and not the integration 
range chosen to form the energy scan from the {\bf Q}-dependent data.  The 
results are summarized in Table~\ref{props}.  The error bars given for the 
values of $\Delta$ and $\hbar\gamma$ are a combination of the errors resulting 
from the \textsc{Reslib} fits as well as the extrapolation described above.

\begin{table}[htbp]
\begin{tabular}{|c|c|d|c|c|}
\hline
~Phase~ & Wavevector & 
\multicolumn{1}{c|}{$\cal A$} & $\Delta$ & $\hbar\gamma$
\\
& & \multicolumn{1}{c|}{~~(barn $\cdot$ meV)~~} & ~(meV)~ & ~(meV)~ \\
\hline
PM & $Z$ & 1.00(8) & 2.3(5) & 2.4(4) \\
PM & $\Sigma$ & 3.0(2) & 2.2(6) & 1.8(2) \\
\hline
HO & $Z$ & 4.3(3) & 2.3(4) &0.9(1) \\
HO & $\Sigma$ & 5.1(3) & 4.2(2) & 0.7(1) \\
\hline
AF & $Z$ & 5.8(6) & 2.3(4) & 0.9(2) \\
AF & $\Sigma$ & 6.1(1.5) & 5.5(3) & 0.7(1)\\
\hline
\end{tabular}
\caption[]{Results of fitting the data in Fig.~\ref{I_vs_hw} to the 
dispersion in Eq.~\ref{eq1}.  The determination of the errors for $\Delta$ and 
$\hbar\gamma$ are described in the text, while the error bars given for ${\cal 
A}$ are a combination of the fitting error and the error from normalization, 
which was 6\%.}
\label{props}
\end{table}

In the HO phase, the excitation at the $\Sigma$ point becomes gapped, with 
$\Delta= 4.2(2)$~meV. Upon entering the AF phase this gap increases to 
$\Delta=5.5(3)$~meV, while the physical half width extracted from this 
analysis, $\hbar\gamma=0.7(1)$~meV, is identical in the two phases. At the 
$Z$ point the gap and width of the spectrum are also identical in the two 
phases. Note that the values for the gap and half width $\Delta=2.3(4)$~meV 
and $\hbar\gamma=0.9(2)$~meV are both larger than literature 
values~\cite{Bourdarot_14} and this may be a result of the coarser ${\bf 
Q}$-resolution of the present measurement. The main difference in the 
scattering in the AF phase as compared to the HO phase is the increased gap at 
the $\Sigma$ and the additional Bragg scattering at (1,0,0). The first moment 
${\cal A}$ at the $Z$ and $\Sigma$ points are within error bars of the values 
in the HO phase, as may also be appreciated by comparing Fig.~\ref{I_vs_hw}(b) 
and Fig.~\ref{I_vs_hw}(c). The inset to Fig.~\ref{I_vs_hw}(c) shows a 
transverse cut through the (1,0,0) elastic peak, the intensity of which 
corresponds to a moment size of 0.36(9)~$\mu_B$.  This is evidence that the 
measurements were indeed conducted in the AF phase. Our observation of 
inelastic scattering at $Z$ is not a surprise given the enhanced AF order. In 
previous, lower pressure work an inelastic peak was observed at $Z$ for 
$P=0.72$~GPa~\cite{Aoki_09} but not for $P=0.62$~GPa~\cite{Bourdarot_10}. A 
possible explanation for all three neutron experiments under pressure is that 
the $Z$ mode softens at the critical pressure and so falls within the elastic 
line in the lower pressure measurements. This would be consistent with recent 
high pressure Raman data~\cite{Blumberg_Private}.

We also note that the ${\bf Q}$-widths of the inelastic magnetic scattering in 
the AF and HO phases are similar and both broader than in the PM 
phase. The limited statistical quality of the AF phase data however, leaves it 
open for now whether or not there are coherent modes in the AF phase as in the 
HO phase. Between the paramagnetic and hidden order phases, transport and 
thermodynamic measurements indicate significant Fermi surface 
reconstruction~\cite{Palstra_85,Maple_86}.  Resistivity~\cite{McElfresh_87}, 
and quantum oscillation measurements~\cite{Hassinger_10b}, on the other hand, 
are much less affected by the HO to AF transition.  Together with the  
similarities between the HO and AF spin correlations reported here, this 
suggests differences between these two phases of URu$_2$Si$_2$ are very subtle.

Apart from inducing or at least enhancing AF order, applied pressure shifts 
$\Sigma$-point intensity to slightly higher energies. This indicates a 
stabilization of AF order under hydrostatic pressure.  Previous work 
interprets gapped excitations at the $Z$ point as a signature of the HO 
phase~\cite{Villaume_08}.  However, the present data shows that entering the 
AF phase does not weaken or destroy either set of excitations.  Likewise, 
pressure does not suppress the HO transition, but actually increases $T_0$, 
before the AF phase emerges~\cite{Motoyama_03}.  All these observations point 
to a significant kinship between the HO and AF phases of URu$_2$Si$_2$.

\begin{acknowledgments}

The authors would like to thank M.B.~Stone for help with the data analysis.  
Work at IQM was supported by DoE, Office of Basic Energy Sciences, Division 
of Materials Sciences and Engineering under award DE-FG02-08ER46544.  This 
work utilized facilities supported in part by the National Science Foundation 
under Agreement No. DMR-1508249.  Research at McMaster University is supported 
by NSERC.  T.J.W. acknowledges support from the Wigner Fellowship program at 
Oak Ridge National Laboratory.

\end{acknowledgments}


\begin{thebibliography}{99}

\bibitem{Palstra_85}T.T.M.~Palstra, A.A.~Menovsky, J.~van~den~Berg, 
A.J.~Dirkmaat, P.H.~Kes, G.J.~Nieuwenhuys and J.A.~Mydosh. Phys. Rev. Lett. 
{\bf 55}, 2727 (1985).

\bibitem{Broholm_87}C.~Broholm, J.K.~Kjems, W.J.L. Buyers, P.~Matthews, 
T.T.M.~Palstra, A.A.~Menovsky and J.A.~Mydosh. Phys. Rev. Lett. {\bf 58}, 1467 
(1987).

\bibitem{Bonn_88}D.A.~Bonn, J.D.~Garrett and T.~Timusk. Phys. Rev. Lett. {\bf 
61}, 1305 (1988).

\bibitem{Buyers_94}W.J.L.~Buyers, Z.~Tun, T.~Peterson, T.E.~Mason,
J.-G.~Lussier, B.D.~Gaulin and A.A.~Menovsky. Physica B {\bf 199\&200}, 95 
(1994).

\bibitem{Mason_95}T.E.~Mason, W.J.L.~Buyers, T.~Peterson, A.A.~Menovsky and 
J.D.~Garrett. J. Phys.: Condens. Matter {\bf 7}, 5089 (1995).

\bibitem{Bourdarot_05}F.~Bourdarot, A.~Bombardi, P.~Burlet, M.~Enderle, 
J.~Flouquet, P.~Lejay, N.Kernavanois, V.P.~Mineev, L.Paolasini, 
M.E.~Zhitomirsky and B.~F$\dot{a}$k. Physica B {\bf 359-361}, 986 (2005).

\bibitem{Bourdarot_14}F.~Bourdarot, S.~Raymond and L.-P.~Regnault. Phil. Mag. 
{\bf 94}, 3702 (2014).

\bibitem{Takagi_07}S.~Takagi, S.~Ishihara, S.~Saitoh, H.-i.~Sasaki, H.~Tanida, 
M.~Yokoyama and H.~Amitsuka. J. Phys. Soc. Japan. {\bf 76}, 033708 (2007).

\bibitem{Wiebe_07}C.R.~Wiebe, J.A.~Janik, G.J.~MacDougall, G.M.~Luke, 
J.D.~Garrett, H.D.~Zhou, Y.-J.~Jo, L.~Balicas, Y.~Qiu, J.R.D.~Copley,
Z.~Yamani and W.J.L.~Buyers. Nature Physics {\bf 3}, 96 (2007).

\bibitem{Aoki_09}D.~Aoki, F.~Bourdarot, E.~Hassinger, G.~Knebel,
A.~Miyake, S.~Raymond, V.~Taufour and J.~Flouquet. J. Phys. Soc. Jap. {\bf 
78}, 053701 (2009).

\bibitem{Butch_10}N.P.~Butch, J.R.~Jeffries, S.~Chi, J.B.~Le\~{a}o, J.W.~Lynn 
and M.B.~Maple. Phys. Rev. B. {\bf 82}, 060408(R) (2010).

\bibitem{Hassinger_10a}E.~Hassinger, D.~Aoki, F.~Bourdarot, G.~Knebel, 
V.~Taufour, S.~Raymond, A.~Villaume and J.~Flouquet. J. Phys.: Conf. 
Ser. {\bf 251}, 012001 (2010).

\bibitem{Bourdarot_10}F.~Bourdarot, E.~Hassinger, S.~Raymond, D.~Aoki,
V.~Taufour, L.-P.~Regnault and
J.~Flouquet. J. Phys. Soc. Jap {\bf 78}, 064719 (2010).

\bibitem{Rodriguez_08}J.A.~Rodriguez, D.M.~Adler, P.C.~Brand, C.~Broholm, 
J.C.~Cook, C.~Brocker, R.~Hammond, Z.~Huang, P.~Hundertmark, J.W.~Lynn, 
N.C.~Maliszewkyj, J.~Moyer, J.~Orndorff, D.~Pierce, T.D.~Pike, 
G.~Scharfstein, S.A.~Smee and R.~Vilaseca. Meas. Sci. Technol. {\bf 19}, 
034023 (2008).

\bibitem{Butch_15}N.P.~Butch, M.E.~Manley, J.R.~Jeffries, M.~Janoschek, 
K.~Huang, M.B.~Maple, A.H.~Said, B.M.~Leu and J.W.~Lynn. Phys. Rev. B. {\bf 
91}, 035128 (2015).

\bibitem{Wiebe_04}C.R.~Wiebe, G.M.~Luke, Z.~Yamani, A.A.~Menovsky and
W.J.L.~Buyers. Phys. Rev. B {\bf 69}, 132418 (2004).

\bibitem{Broholm_91}C.~Broholm, H.~Lin, P.T.~Matthews, T.E.~Mason,
W.J.L.~Buyers, M.F.~Collins, A.A.~Menovsky, J.A.~Mydosh and
J.K.~Kjems. Phys. Rev. B {\bf 43}, 12809 (1991).

\bibitem{Zheludev_07}A.~Zheludev: ResLib 3.4 (Oak Ridge National     
Laboratory, (2007).

\bibitem{Blumberg_Private}G.~Blumberg. Private communication.

\bibitem{Maple_86}M.B.~Maple, J.W.~Chen, Y.~Dalichaouch, T.~Kohara, 
C.~Rossel, M.S.~Torikachvili, M.W.~McElfresh and J.D.~Thompson.  Phys. 
Rev. Lett. {\bf 56}, 185 (1986).

\bibitem{McElfresh_87}M.W.~McElfresh, J.D.~Thompson, J.O.~Willis, M.B.~Maple, 
T.~Kohara and M.S.~Torikachvili. Phys. Rev. B {\bf 35}, 43 (1987).

\bibitem{Hassinger_10b}E.~Hassinger, G.~Knebel, T.D.~Matsuda, D.~Aoki, 
V.~Taufour and J.~Flouquet. Phys. Rev. Lett. {\bf 105}, 216409 (2010).

\bibitem{Villaume_08}A.~Villaume, F.~Bourdarot, E.~Hassinger,
S.~Raymond, V.~Taufour, D.~Aoki and J.~Flouquet. Phys. Rev. B {\bf 78}, 012504 
(2008).

\bibitem{Motoyama_03}G.~Motoyama, T. Nishioka and N.K.~Sato. Phys. Rev. 
Lett. {\bf 90}, 166402 (2003).

\end{thebibliography}
\end{document}